\documentclass[11pt,twoside]{article}

\usepackage{aspsoho23}
\usepackage{epsf}
\usepackage{lscape}
\usepackage{graphicx}
\begin{document}
\title{The solar wind as seen by SOHO/SWAN since 1996: comparison with SOHO/LASCO 
          C2 coronal densities}   %%% Fill in title
\author{Lallement R.$^{1}$, Qu\'emerais E.$^{1}$, Lamy P.$^{2}$ , Bertaux J.L.$^{1}$, Ferron S.$^{3}$, Schmidt W.$^{4}$}   %%% Fill in author names
\affil{$^{1}$ LATMOS-IPSL, CNRS, Universit\'e Versailles-Saint-Quentin, France, $^{2}$LAM Marseille, France, $^{3}$ACRI-ST, Sofia-Antipolis, France, $^{4}$FMI, Helsinki, Finland}    %%% Fill in author affiliations

\begin{abstract} %%% Abstract to run on from here.
We update the SOHO/SWAN H Lyman-$\alpha$ brightness analysis to cover the 1996-2008 time interval.  A forward model applied to the intensity maps provides the latitude and time dependence of the interstellar Hydrogen ionisation rate over more than a full solar cycle. The hydrogen ionisation, being almost entirely due to charge-exchange with solar wind ions, reflects closely the solar wind flux. Our results show that the solar wind latitudinal structure during the present solar minimum is strikingly different from the previous minimum, with a much wider slow solar wind equatorial belt which persists until at least the end of 2008.  \\
We compute absolute values of the in-ecliptic  H ionisation rates using OMNI solar wind data and use them to calibrate our ionisation rates at all heliographic latitudes. We then compare the resulting fluxes with the synoptic LASCO/C2 electron densities at 6 solar radii. The two time-latitude patterns are strikingly similar over all the cycle.  This comparison shows that  6R$_{s}$ densities can be used to infer the solar wind type close to its source, with high (resp. low) densities tracing the slow (resp. fast)  solar wind, simply because the density reflects at which altitude occurs the acceleration. \\
The comparison between the two minima suggests that the fast polar wind acceleration occurs at larger distance during the current minimum compared to the previous one. This difference, potentially linked to the magnetic field decrease or(and) the coronal temperature decrease should be reproduced  by solar wind expansion models . 
\end{abstract}

%%% MAIN BODY OF TEXT GOES HERE. CONSULT "INSTRUCTIONS FOR AUTHORS USING
%%% LATEX2E MARKUP", SECTIONS 2.3-2.6 FOR HELP WITH EQUATIONS, FIGURES,
%%% AND TABLES.

\begin{figure}
\includegraphics[width=14cm,height=6cm]{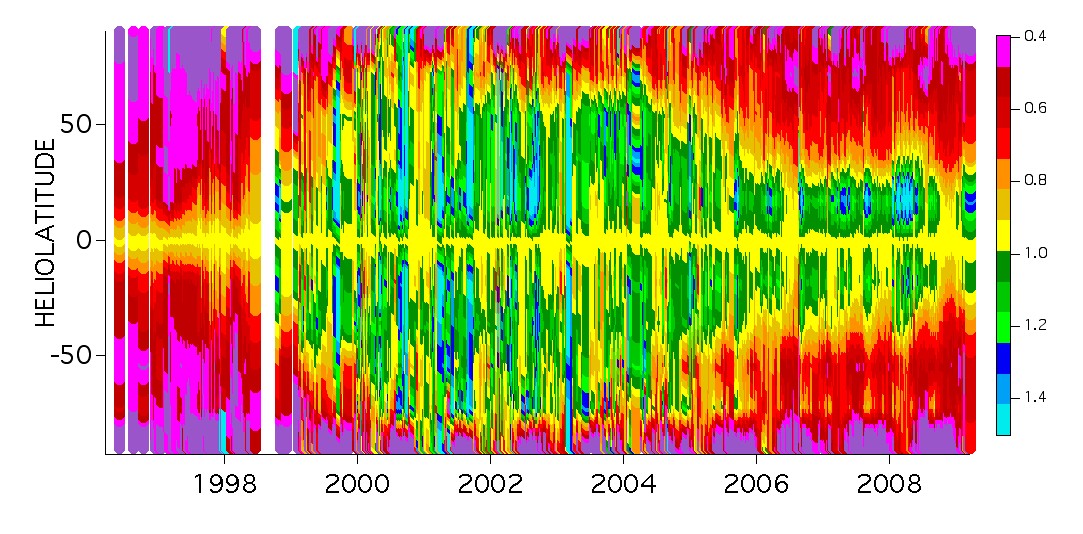}
  \caption{H ionisation rates from SWAN brightness maps. Ionisation rates have been normalised to the equator sector. Note that the polar rate is more poorly defined due to the shorter time atoms spend traveling above the solar pole, which explains the high variability in this area. There is $\textbf{a very strong difference between the two minima}$, with a much wider slow solar wind (high ionisation) equatorial belt during the present minimum.}
     \label{rates}
\end{figure}

\section{SWAN maps and interstellar H retrieval} 

The SWAN instrument on board SOHO is recording full-sky maps of the diffuse Ly-alpha background emission with a resolution of 1x1 deg$^{2}$. For details about the instrument and the observing strategies see e.g. \citet{bert95}, \citet{quem99}, \citet{cost99}. 
Except for a contamination by the bright early-type stars, which is suppressed by means of a mask in the data processing, all the signal is the solar Lyman-alpha radiation backscattered by interstellar H atoms which are permanently flowing within the heliosphere. The distribution of the H atoms is deeply affected by the solar wind (SW) through charge transfer with solar ions and also photo-ionization, which allows to infer solar wind fluxes as a function of time and heliolatitude. Here we update the solar wind large scale distribution derived from SWAN brightness maps. 
The H distribution and therefore the sky Ly-$\alpha$ brightness pattern reflects the ionization of the neutral gas by the solar wind (the main effect) and the radiation. A forward model allowing for a latitude-dependent ionization has been adjusted to the SWAN maps.  For details about the model and the method see \citet{lall85}, \citet{quem06}. In short, the 3D space around the sun is divided in heliographic latitude bins, each of those being characterised by a given 1 AU ionisation rate (the ionisation varies everywhere as r$^{-2}$). Every representative H atom is followed along its trajectory and its ionisation at a given time t  is computed according to its latitude at this time. A least-squares adjustment of the 1 AU ionisation in each latitude bin is performed on each map. Fig. \ref{rates} displays the results from 830 maps between 1996 and 2009. The quantity shown is the H global ionization rate $\tau$(H) at 1AU, here normalised to the solar equator sector. The marked difference between the minimum and maximum solar phases is the dominant pattern. High ionisation rates occur at low heliographic latitudes at solar min, and correspond to the slow and dense solar wind, while low rates correspond to the fast polar wind. At maximum the solar wind is more isotropic, as does the ionisation pattern.  The ionisation rate (see below) is doubly affected by the solar wind type. The fast wind has a lower flux, and the ionisation efficiency is also smaller due to the reduced charge-transfer cross-section. \\
There is also a striking difference between the two minima. While in 1996-97 the slow wind low-latitude belt is 25$\deg$ wide only, the rest being occupied by low flux high speed wind, there is a much wider slow wind belt (up to 90-100$\deg$) in 2006-2008. This is in agreement with Ulysses electron and ion data and IPS speed latitudinal structure (\citet{issau08}, \citet{ebert09}, \citet{mccom08}, \citet{toku09}), but here we additionally show that this is a persisting feature, reflected in the global pattern until at least 2008.\\

%\begin{figure}
\begin{figure}
\includegraphics[height=6cm]{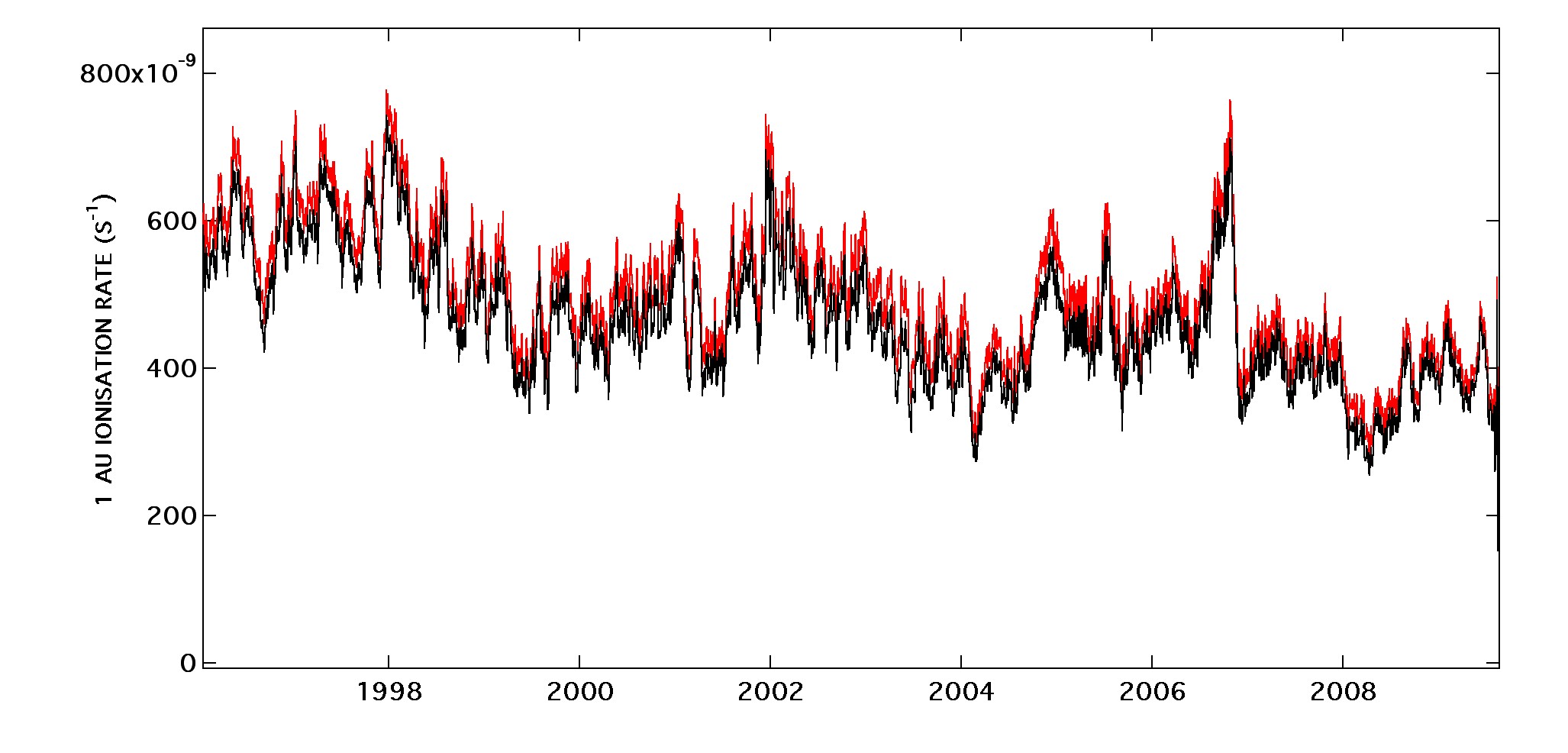}
% \plotone{OMNICX.epsf}
  \caption{In-ecliptic H ionisation rates due to charge-transfer with solar ions, based on OMNI data. The two curves correspond to the two available empirical formulae for the charge transfer cross-section dependence on velocity. }
 \label{omnicx}
\end{figure}

%\begin{landscape}
\begin{figure}
\includegraphics[height=6cm]{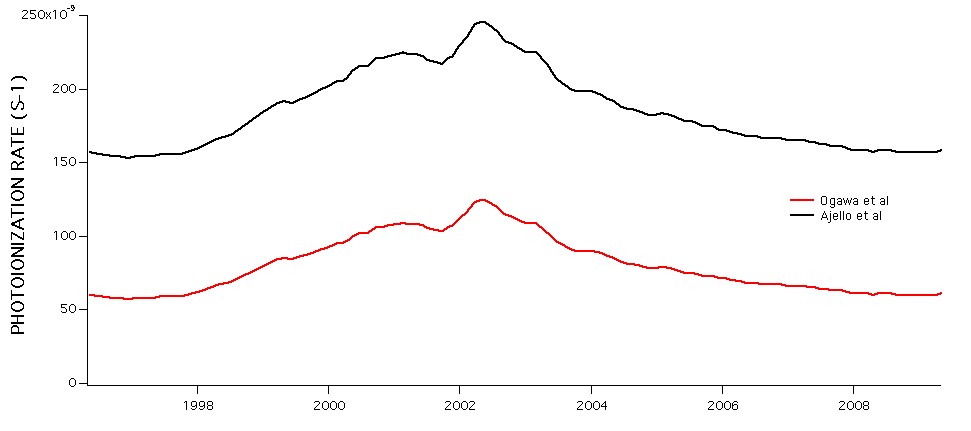}
  \caption{H photo-ionisation rates based on the formulae of \citet{ajello87} and \citet{ogawa95}, and on the composite Ly-$\alpha$ intensities from \citet{woods00}. }
     \label{phot}
\end{figure}

\section{In-ecliptic calibration and absolute values of the ionisation rates over the cycle}
\begin{figure}
\includegraphics[width=16cm,height=6cm]{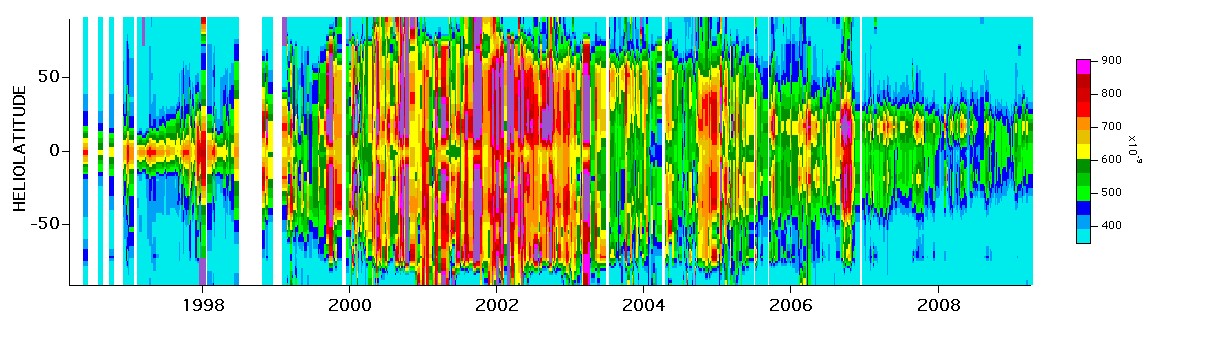}
  \caption{Absolute H ionisation rates from SWAN, scaled after in-ecliptic values.}
     \label{betascaled}
\end{figure}

The H ionisation is the sum of the ionization by charge-transfer with the solar wind ions (mostly protons) and, in much less proportion, the photo-ionization (10-20\% at maximum). The  former term is the product of the solar wind flux by the charge transfer cross-section $\sigma$, which has a measured dependence on the velocity. Since the cross-section is a decreasing function of the collisional energy, the high speed solar wind speed  has a weaker charge-transfer rate. Because the high speed wind is also generally characterised by a lower flux, the two effects conjugate themselves, resulting in the significant decrease of the rate we are observing at high latitude. \\
In order to avoid any potential influence of the instrument calibration (work based on multi-spacecraft data is in progress), we have calibrated our ionisation on the equatorial region by using in-ecliptic  data from the NASA-GSFC OMNI database. Using the high resolution data (1mn) of solar wind speeds and densities we have computed the in-ecliptic H charge-transfer ionisation rate and a 1 month running mean average. The results are shown in Fig \ref{omnicx} for the two existing different velocity dependence empirical formulations of the cross-section. We have considered these rates as appropriate for the equatorial  region, assuming that the solar rotation and the 7$\deg$ solar equator inclination result in a smoothing of those rates to give a zero heliographic latitude monthly averaged value. 
 We then have used two proxies for the photo-ionisation rate as a function of time, based on the composite Ly-$\alpha$ irradiance  of \citet{woods00} and the photo-ionisation rates of \citet{ajello87} and \citet{ogawa95}, assuming that the photo-ionisation is an increasing function of the Ly-$\alpha$ intensity. The results are shown in Fig \ref{phot} . These two rates are significantly different in absolute values and we have intentionally used both of them to test the influence of uncertainties on the photo-ionisation on our results. We have added the photo-ionisation contribution to the SW ionisation to give the total ionisation rate. Finally, all our equator-normalised ionisation rates have been multiplied by these low-latitude rates. 
 The resulting absolute values of the H ionisation are displayed in Fig \ref{betascaled} for the lower photo-ionisation rate. The overall pattern would be very similar and absolute values slightly higher in the case of the higher rate. The last step, not shown here, is the conversion into solar wind fluxes and is the subject of ongoing work (making use of IPS data in order to take into account the cross-section variations). However, the final flux pattern will be very similar to the global rate pattern shown here. The main characteristics described in the previous section remain, namely the strong minimum-maximum difference in latitudinal extent and rate absolute value, as well as the strong difference between the two consecutive minima.

\begin{figure}
\includegraphics[width=16cm,height=6cm]{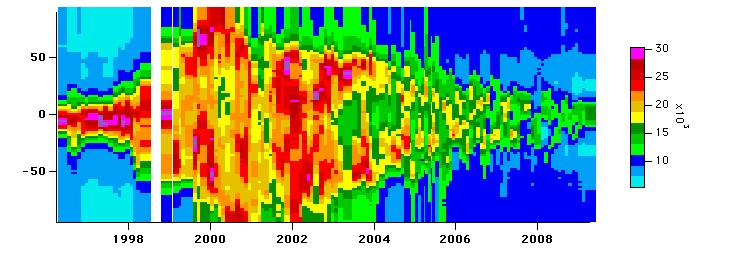}
  \caption{LASCO C2 electron densities at 6 solar radii (see \citet{quem07} for more details about the analysis). There is $\textbf {a striking similarity with the  time-latitude map of SWAN ionisation rates}$ shown in Fig \ref{betascaled}. Interestingly, the high latitude density is larger during the present minimum compared to the previous one. Because the measured solar wind fluxes are significantly smaller (see text), this implies a different expansion of the wind, namely an acceleration at a higher altitude.}
     \label{lasco6rs}
\end{figure}

\section{Combination with LASCO density profiles}
Fig \ref{lasco6rs} displays the LASCO C2 average electron density at 6 R$_{S}$. There is obviously  a very strong similarity between the H ionisation pattern and the 6 Rs electron density. This has already been discussed by \citet{quem07}. The time-latitude distribution of the high electron densities follows very closely the SWAN high ionisation rate distribution. This is interpreted as due to the strong correspondence between  fast speed wind (which produces low H ionisation) and low density in its source regions.\\
We want to emphasise here the potential interest of combining these data sets, as well as speed measurements from IPS (e.g. \citet{toku09}). The knowledge of the particle flux and the speed at large distance, coupled to the density profile within the acceleration region allows the derivation of the mean speed profile, simply from mass conservation along flow tubes. Indeed, the combination of the SWAN SW fluxes deduced from the ionization rates and the LASCO data between 2.5 and 7 Rs has led to interesting results for the 1996 minimum (\citet{quem07}. These authors have combined the three different data and found that the fast solar wind has almost reached its final speed at 6 Rs, while at the same altitude the slow wind has only reached about half of its final speed. This behaviour had been inferred in some particular cases, but the SWAN/LASCO/IPS combination demonstrates that it is a general property of the solar wind, valid at any time and at all latitudes. It is important to note that the low 6Rs density for the fast wind is not solely due to a reduced particle flux, it is also due to the low altitude at which the solar wind is accelerated and escapes. Such density differences would occur even in the case of a uniform particle flux and different acceleration altitudes. 

\section{Different accelerations at the two solar minima?}
Such a combined study of SWAN, LASCO and IPS data is still in progress for the present minimum of activity.  From now on however it is possible to foresee a peculiar result. As a matter of fact, the LASCO map (\ref{lasco6rs}) shows that the electron density above 40¡ latitude is higher during this minimum than in 96 (note also that there is a mid-latitude minimum, not present in 96). On the other hand, it is clear from a number of data that the high-latitude SW fluxes are 20-25\% below the previous minimum fluxes (\citet{issau08}, \citet{ebert09}, \citet{mccom08}). This is confirmed by SWAN, using the normalized values of Figure \ref{betascaled}. Finally, the terminal speeds are similar, or only a few \% less, as shown by SW data and IPS (\citet{toku09}). If the coronal velocity profiles were unchanged, lower fluxes and identical final speeds would imply lower densities at every altitude, which is contradicted by LASCO. This suggests that the velocity profile at high latitude is different from what it has been at the last minimum, with less acceleration (leading to higher densities) at 6-7 Rs. That the acceleration of the fast wind occurs at a higher altitude (above 6 Rs) in polar holes, in conjunction with lower coronal magnetic fields, SW fluxes and temperatures is an interesting test for solar wind expansion models .

 %\subsection{}   %%% Second level section head (remove "%" symbol)
%\subsubsection{}   %%% Lowest level section head (remove "%" symbol)

%%% Unnumbered top level section head (remove "%" symbol)
%\subsection*{}   %%% Unnumbered second level section head (remove "%" symbol)

%\end{landscape}

\acknowledgements %%% Text of acknowledgements runs on after this command.
SOHO is a mission of international cooperation between ESA and NASA. The SWAN analysis is financed in France by CNES with support from CNRS and in Finland by TEKES and the Finnish Meteorological Institute. We have made extensive use of the NASA-GSFC OMNIweb Service.

%%% THE BIBLIOGRAPHY
%%%
%%% CONSULT SECTION 3 OF "INSTRUCTIONS FOR AUTHORS" FOR HOW TO USE NATBIB.
%%% AUTHORS ARE ENCOURAGED TO USE EITHER THE "THEBIBLIOGRAPY" ENVIRONMENT
%%% BY UNCOMMENTING (DELETING THE "%" SYMBOL) THE COMMANDS BELOW, OR BY
%%% USING THE BIBTEX ENVIRONMENT. TO FIND OUT WHICH IS APPLICABLE TO YOUR
%%% CONTRIBUTION, CONSULT THE VOLUME EDITORS FOR YOUR PROCEEDINGS.
%%%

\end{document}